\documentclass{iaus}
\usepackage{graphicx}

\title[IAUS 292.~~Supershells \& Molecular Cloud Formation] 
{The Supershell-Molecular Cloud Connection in the Milky Way and Beyond}

\author[J. R. Dawson]   
{J. R. Dawson$^1$, N. M. McClure-Griffiths$^2$, Y. Fukui$^3$, J. Dickey$^1$, T. Wong$^4$, A. Hughes$^5$, A. Kawamura$^6$}

\affiliation{$^1$School of Mathematics \& Physics, University of Tasmania, Sandy Bay, TAS 7005, Australia \\ email: {\tt joanne.dawson@utas.edu.au} \\[\affilskip] $^2$Australia Telescope National Facility, CASS, Marsfield NSW 2122, Australia \\[\affilskip] $^3$Dept. of Astrophysics, Nagoya University, Chikusa-ku, Nagoya, Japan \\[\affilskip] $^4$Astronomy Department, University of Illinois, Urbana, IL 61801, USA  \\[\affilskip]$^5$Max-Planck-Institut f\"ur Astronomie, K\"onigstuhl 17, D-69117, Heidelberg, Germany \\[\affilskip]$^6$National Astronomical Observatory of Japan, Tokyo 181-8588, Japan
}
\pubyear{2013}
\volume{292}  
\pagerange{000}
\jname{Molecular Gas, Dust, and Star Formation in Galaxies}
\editors{T. Wong \& J. Ott, eds.}
\begin{document}

\maketitle

\begin{abstract}
The role of large-scale stellar feedback in the formation of molecular clouds has been investigated observationally by examining the relationship between H{\sc i} and $^{12}$CO(J=1--0) in supershells. Detailed parsec-resolution case studies of two Milky Way supershells demonstrate an enhanced level of molecularisation over both objects, and hence provide the first quantitative observational evidence of increased molecular cloud production in volumes of space affected by supershell activity. Recent results on supergiant shells in the LMC suggest that while they do indeed help to organise the ISM into over-dense structures, their global contribution to molecular cloud formation is of the order of only $\sim10\%$. 
\end{abstract}

\firstsection 
\section{Introduction}

The formation of cold, dense molecular gas from the atomic interstellar medium (ISM) is the first stage of the star formation process, and sets fundamental constraints on a galaxy's star formation rate. 
Astrophysical drivers of molecular cloud formation include global gravitational instabilities \cite[(e.g. Tasker \& Tan 2009)]{tasker09}, the accumulation of matter in spiral shocks \cite[(e.g. Dobbs \& Bonnell 2008)]{dobbs08}, and compression in large-scale expanding shells driven by stellar feedback \cite[(e.g. Ntormousi et al. 2011)]{ntormousi11}; in all cases aided by turbulent compression that enhances density on a range of scales \cite[(Glover \& Mac Low 2007)]{glover07}. However, disentangling the relative contributions of these processes is usually not trivial, and the primary drivers are still open to debate. 

In this work we focus on the formation of molecular clouds in expanding supershells created by the stellar feedback from OB clusters. The theoretical context for this is the compression, cooling and fragmentation of the atomic medium in turbulent shocks and flows, of which shells swept-up by large-scale stellar feedback are but one example \cite[(see review by Vazquez-Semadeni 2010)]{vazquez10}. However, 
while there is some suggestion from numerical models that the contribution of stellar feedback to molecular gas production may be small \cite[(e.g. Joung \& Mac Low 2006)]{joung06}, very little has been observationally constrained. We need to answer the question, ``what fraction of a galaxy's molecular gas is formed by stellar feedback?'' 

\section{Milky Way Supershells}

In the Milky Way, detailed parsec-resolution studies of $^{12}$CO(J=1--0) and H{\sc i} have been carried out in the Galactic supershells GSH287+04--17 and GSH277+00+36, which are both large ($R\sim150$--$300$ pc), gently expanding ($v_{exp}\sim10$--$20$ km s$^{-1}$) and highly evolved ($t\sim10^7$ yr) `chimney' systems \cite[(Dawson et al. 2011a)]{dawson11a} . 
Figure 1 shows a section of the wall of GSH 287+04--17, showcasing features commonly seen in the walls of both shells. Features labelled \textit{a} indicate small ($M_{\mathrm{H}_2}\sim10^{2\mathrm{-}3} M_{\odot}$) clumps of molecular gas offset towards the tips of atomic `fingers' that point in the direction of the shell centre. 
These clouds have no substantial H{\sc i} envelopes and no evidence for a hidden `dark gas' component (either opaque H{\sc i} or CO-dark H$_2$), with some also showing shock-disrupted morphology \cite[(Dawson et al. 2011b)]{dawson11b}. 
These objects are likely pre-existing clouds being disrupted by their interaction with the shell, with the lack of material to shield CO implying maximum survival times of a few Myr. 

\begin{figure}[t]
\begin{center}
 \includegraphics[width=3.0in]{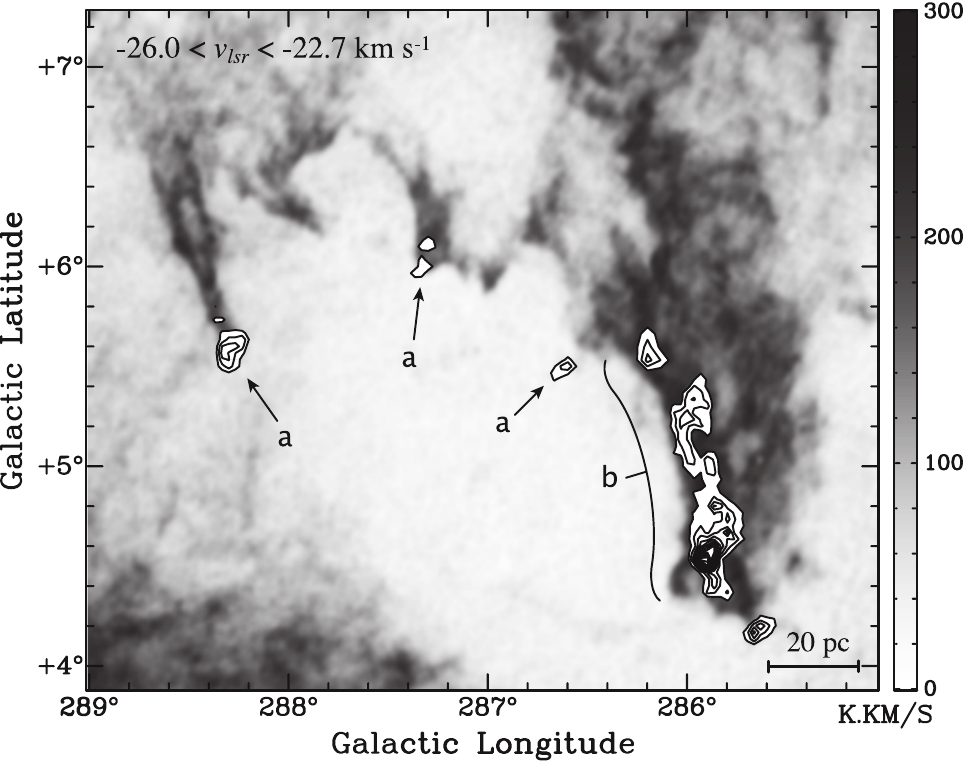} 
 \caption{Subsection of the wall of GSH 287+04--17. The greyscale is H{\sc i} and the filled contours are $^{12}$CO(J=1--0). Features labelled 'a' indicate CO clouds that are likely being destroyed. The feature labelled 'b' shows a cloud complex that is likely being formed in the shell wall. 
 }
   \label{fig1}
\end{center}
\end{figure}

The feature labelled \textit{b} is an example of a larger ($M_{\mathrm{H}_2}\sim10^{4} M_{\odot}$) molecular cloud that is well embedded in atomic material, and forms a coherent part of the main shell wall. This cloud shows evidence for a substantial dark component identified from $100\mu$m IR excess, which comprises over 50\% of the total mass of the H{\sc i}-CO complex, and provides sufficient material to 
shield CO molecules against UV dissociation \cite[(Dawson 2011a)]{dawson11a}. The initial number densities and timescales are consistent with a scenario in which the molecular gas has been formed in-situ over the lifetime of the shell. 

\begin{figure}[t]
\begin{center}
 \includegraphics[width=3.8in]{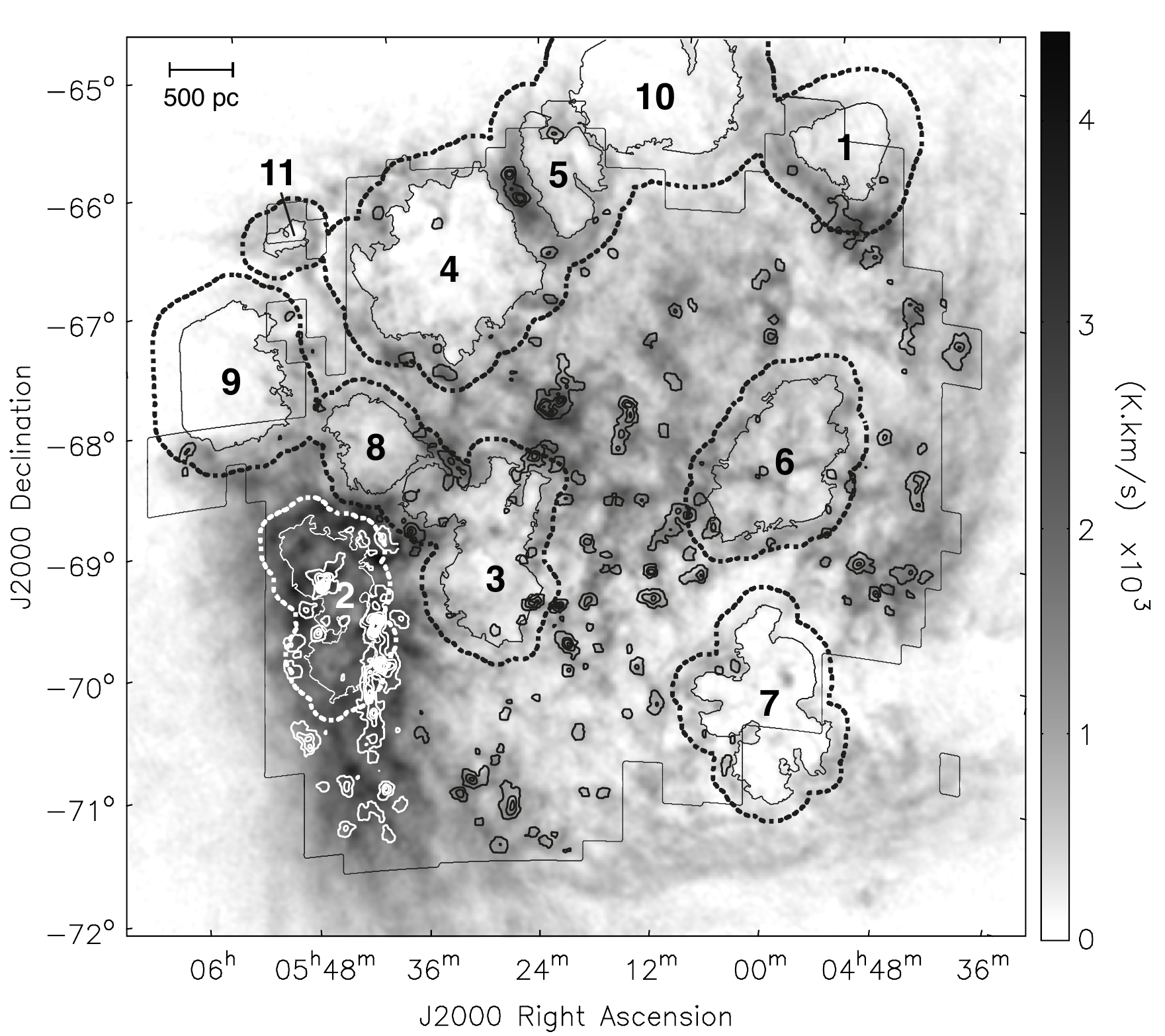} 
 \caption{Figure 2 from \cite{dawson12}, showing SGSs in the LMC. The greyscale shows H{\sc i} \cite[(Kim et al. 2003)]{kim03} and the contours show $^{12}$CO(J=1--0) \cite[(Fukui et al. 2008)]{fukui08}. Solid lines show the thresholded inner rims of supergiant shell complexes and dotted lines show the outer rims (delineating the outer edges of the dense shells).}
   \label{fig2}
\end{center}
\end{figure}

From a galaxy evolution perspective, the most important information is the \textit{net effect} that supershells have on the quantity of molecular gas in the volumes they occupy. 
This may be estimated by comparing the molecular fraction, $f_{\mathrm{H}_2} = M_{\mathrm{H}_2} / (M_{\mathrm{HI}} + M_{\mathrm{H}_2})$, in shell volumes -- including the evacuated voids -- to that in 
their local vicinities (essentially a proxy for the undisturbed medium). Since $f_{\mathrm{H}_2}$ varies with 
location in the Galactic Disk, it is important that these `background' regions are restricted carefully in $l$-$b$-$v$ space to include only emission that is genuinely local to the shells. 
The results of this analysis reveal that $f_{\mathrm{H}_2}$ in the supershell volumes is enhanced by a factor of $\sim2$--3 with respect to their local backgrounds, implying that as much as $50$--$70\%$ of the molecular matter in their walls may have been formed as a direct result of stellar feedback. These results are discussed in detail in Dawson et al. (2011a).

\section{Supergiant Shells in the LMC} 

The above results for Galactic objects are excellent proof of concept. However it is unclear whether these two systems are representative of the general case, either in the Milky Way or elsewhere. For a more statistical approach we turn to the Large Magellanic Cloud (LMC).  

Figure \ref{fig2} shows the outlines of supergiant shells (or in some cases complexes of nested or overlapping shells) in the LMC. Supergiant shells (SGSs) are the extreme end of the supershell population, defined as those objects whose radii are larger than the neutral gas scale height ($\sim180$ pc), meaning that they are assumed to have blown out of both sides of the disk. We select all supergiant shells from existing H{\sc i} and H$\alpha$ catalogues \cite[(Meaburn 1980, Kim et al. 1999)]{meaburn80,kim99}, with likely false detections excluded \cite[(Book et al. 2008)]{book08}, and define their shape and extent accurately by thresholding in H{\sc i} channel maps. Further details of the SGS definition method are given in \cite[Dawson et al. (2012)]{dawson12}. 

If large-scale stellar feedback was the dominant method of molecular cloud production in the LMC, we would expect to see a significant enhancement in $f_{\mathrm{H}_2}$ in SGS volumes compared to the rest of the LMC disk. This is not the case. The molecular fraction in all zones occupied by SGSs is $0.055\pm0.001$, compared to $0.054\pm0.001$ in the remainder of the disk. The difference between these numbers is surprisingly small, and implies that supergiant shells have a negligible effect on the molecular gas fraction of the LMC. In fact, the closeness is somewhat serendipitous, arising from the fact that extended areas of high molecular fraction are equally sampled by both SGS and background zones. Nevertheless, they underscore the fact that SGSs are clearly not the dominant factor determining where molecular material is formed in the LMC.

\begin{figure}[t]
\begin{center}
 \includegraphics[width=3.4in]{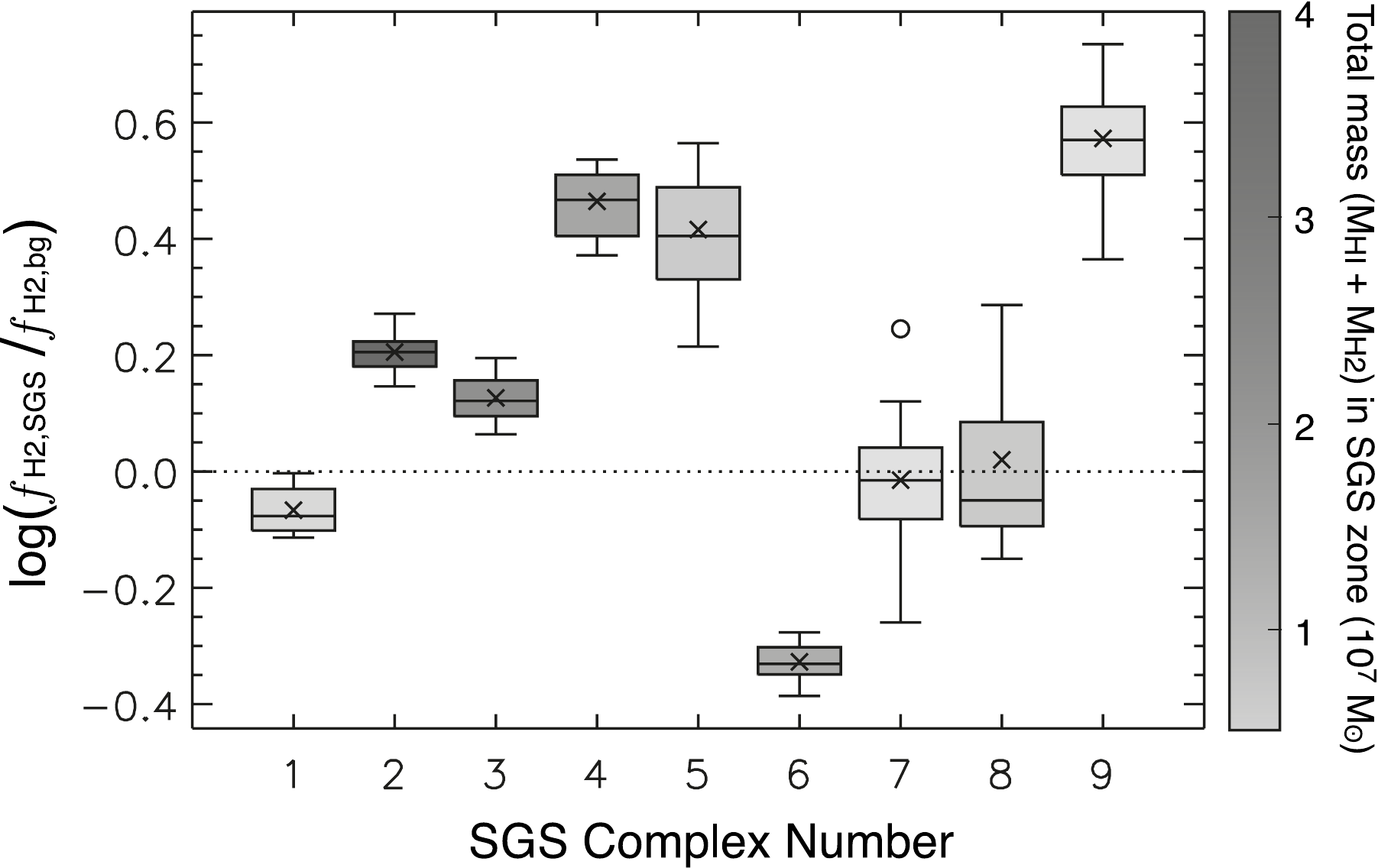} 
 \caption{Figure 4 from \cite[Dawson et al. (2012)]{dawson12} comparing the molecular fraction in individual SGSs, $f_{\mathrm{H}_2,\mathrm{SGS}}$, with that in unrelated material local to the shells, $f_{\mathrm{H}_2,\mathrm{bg}}$. The numbers on the horizontal axis refer to the labels in figure \ref{fig2}. Box plots illustrate the distribution of values obtained for different background zone widths (see Dawson et al. 2012). The dotted line marks the point at which the molecular fractions in the shell and background zones are equal.}
   \label{fig3}
\end{center}
\end{figure}

As in the Milky Way, comparing individual SGSs with their \textit{local} vicinities is a means of minimising the effect of differences in $f_{\mathrm{H}_2}$ that occur due to unrelated environmental factors, and more accurately assessing the effect of a shell. Here, local background zones are defined as bands of constant width around SGSs. 
The results of this analysis are shown in figure \ref{fig3}. Five out of the nine SGS complexes ($\sim70\%$ by mass) show robust evidence of enhanced molecular fractions in shell volumes. While there is considerable scatter over the sample, these results suggest that globally supergiant shells \textit{do} have a measurable effect on the molecular gas fraction of the LMC, albeit a small one. Averaged over the whole sample, the numbers imply that $\sim14$--$25\%$ of the CO in SGSs -- $\sim6$--10$\%$ of the total molecular matter in the LMC -- may have been formed as a result of conversion in the shell walls. 
This is consistent with a picture in which feedback from massive stars plays an important role in structuring the disk and triggering star formation on local scales, but does not drive the initial production of molecular gas on galactic scales. 

\end{document}